\documentclass[a4paper,11pt]{article}

\pdfoutput=1

\usepackage{jcappub}
\usepackage{amsmath}
\usepackage{graphicx}
\usepackage[normalem]{ulem}

\usepackage{physics}
\usepackage{pgfplots}

\usepackage[english]{babel}
\usepackage[utf8]{inputenc}
\usepackage{pdflscape}
\usepackage{enumerate}
\usepackage{amsbsy}
\usepackage{amsmath} 
\usepackage{graphics}
\usepackage{mathrsfs}
\usepackage{wrapfig}
\usepackage{mathtools}

\usepackage{amsfonts}
\usepackage{pstricks}
\usepackage{color}
\usepackage{setspace}
\usepackage{tensor}

\newcommand{\bea}{\begin{eqnarray}} \newcommand{\eea}{\end{eqnarray}}
\newcommand{\el}{\nonumber \\}
\newcommand{\re}[1]{(\ref{#1})}

\newcommand{\pat}{\partial}
\renewcommand{\sec}[1]{section \ref{#1}}
\newcommand{\fig}[1]{figure \ref{#1}}

\newcommand{\para}{\paragraph}

\newcommand{\PR}{\mathcal{P}_\mathcal{R}}

\renewcommand{\a}{\alpha}
\renewcommand{\b}{\beta}
\renewcommand{\c}{\gamma}
\renewcommand{\d}{\delta}

\renewcommand{\l}{\lambda}

\newcommand{\rmd}{\mathrm{d}}

\newcommand{\ie}{i.e.\ }
\newcommand{\eg}{e.g.\ }

\newcommand{\Mpl}{M_{{}_{\mathrm{Pl}}}}

\def\beq{\begin{equation}}
\def\eeq{\end{equation}}
\def\baq{\begin{eqnarray}}
\def\eaq{\end{eqnarray}}

\title{Critical point Higgs inflation in the Palatini formulation}

\author[a,b]{Vera-Maria~Enckell,}
\author[b,c]{Sami Nurmi,}
\author[b,d]{Syksy R\"{a}s\"{a}nen}
\author[e]{and Eemeli Tomberg}

\affiliation[a]{Department of Medical Imaging and Radiation Therapy, Kymenlaakso Central Hospital, Kymenlaakso Social and Health Services (Kymsote), Kotka, Finland}

\affiliation[b]{Helsinki Institute of Physics (HIP),
P.O. Box 64, FIN-00014 University of Helsinki,  Finland}

\affiliation[c]{Department of Physics, University of Jyv\"{a}skyl\"{a}, P.O. Box 35, FIN-40014 University of
Jyv\"{a}skyl\"{a}, Finland}

\affiliation[d]{University of Helsinki, Department of Physics,
P.O. Box 64, FIN-00014 University of Helsinki, Finland}

\affiliation[e]{Laboratory of High Energy and Computational Physics, National Institute of Chemical Physics and Biophysics, R\"avala pst.~10, 10143 Tallinn, Estonia}

\emailAdd{verkkue@gmail.com}
\emailAdd{sami.t.nurmi@jyu.fi}
\emailAdd{syksy.rasanen@iki.fi}
\emailAdd{eemeli.tomberg@kbfi.ee}

\abstract{
We study Higgs inflation in the Palatini formulation with the  renormalisation group improved potential in the case when loop corrections generate a feature similar to an inflection point. Assuming that there is a threshold correction for the Higgs quartic coupling $\l$ and the top Yukawa coupling $y_t$, we scan the three-dimensional parameter space formed by the two jumps and the non-minimal coupling $\xi$.

The spectral index $n_s$ can take any value in the observationally allowed range. The lower limit for the running is $\alpha_s>-3.5\times10^{-3}$, and $\a_s$ can be as large as the observational upper limit. Running of the running is small. The tensor-to-scalar ratio is $2.2\times 10^{-17}<r<2\times10^{-5}$. We find that slow-roll can be violated near the feature, and a possible period of ultra-slow-roll contributes to the widening of the range of CMB predictions. Nevertheless, for the simplest tree-level action, the Palatini formulation remains distinguishable from the metric formulation even when quantum corrections are taken into account, because of the small tensor-to-scalar ratio.
}

\begin{document}

\begin{flushleft}
	\hfill		 HIP-2020-33/TH \\
\end{flushleft}

\maketitle
  
\setcounter{tocdepth}{2}

\setcounter{secnumdepth}{3}

\section{Introduction} \label{sec:intro}

\para{Higgs inflation and its complications.}

Higgs inflation is a conservative model that uses only the degrees of freedom in the Standard Model (SM) of particle physics and general relativity (GR) \cite{Bezrukov:2007} (for reviews, see \cite{Bezrukov:2013, Bezrukov:2015, Rubio:2018}). When the Higgs is directly coupled to the Ricci scalar, the effective Higgs potential is exponentially flat, giving (at tree-level and in the metric formulation of GR) a unique prediction for the spectral index and tensor-to-scalar ratio that is in good agreement with observations \cite{Planck2018}. There are two complications. Quantum corrections can change the potential \cite{Espinosa:2007qp, Barvinsky:2008ia, Barvinsky:2009fy, Barbon:2009, Burgess:2009, Popa:2010xc, DeSimone:2008ei, Bezrukov:2008ej, Bezrukov:2009db, Barvinsky:2009ii, Bezrukov:2010, Bezrukov:2012sa, Allison:2013uaa, Salvio:2013rja, Shaposhnikov:2013ira, Weenink:2010, Calmet:2012eq, Steinwachs:2013tr, Bezrukov:2014bra, Hamada:2014iga, Prokopec:2014, Kamenshchik:2014waa, Burns:2016ric, Fumagalli:2016lls, Hamada:2016, Karamitsos:2017elm, Karamitsos:2018lur, George:2013iia, Postma:2014vaa, Prokopec:2012, Herrero-Valea:2016jzz, Pandey:2016jmv, Pandey:2016unk, Bezrukov:2014ipa, Rubio:2015zia, Enckell:2016, Bezrukov:2017dyv, Rasanen:2017, Masina:2018, Salvio:2017oyf, Ezquiaga:2017fvi, Rasanen:2018a, Enckell:2018a, Shaposhnikov:2020fdv, Fumagalli:2020ody}, and the choice of the formulation of GR can modify the relation between the particle physics potential and the evolution of the inflaton and spacetime \cite{Bauer:2008, Raatikainen:2019}.
 
Quantum corrections affect the mapping between the electroweak (EW) scale parameters and the inflationary observables on the plateau. Part of this is captured by the renormalisation group (RG) running of the Higgs quartic coupling $\l$ on the inflationary plateau, where its value can be quite different from the EW scale value. In most of the parameter space, quantum corrections to the effective Higgs potential are not expected to generate significant changes to the inflationary dynamics \cite{Fumagalli:2016lls}. Nevertheless, in tuned regions of parameter space where the value of $\l$ is of the same order as its running, loop corrections can open new inflationary regimes, like inflection point inflation \cite{Allison:2013uaa, Bezrukov:2014bra, Hamada:2014iga, Bezrukov:2014ipa, Rubio:2015zia, Fumagalli:2016lls, Enckell:2016, Bezrukov:2017dyv, Rasanen:2017, Masina:2018, Salvio:2017oyf, Ezquiaga:2017fvi, Rasanen:2018a} (also called critical point inflation in the case when the potential does not necessarily have an exact inflection point), hilltop inflation \cite{Fumagalli:2016lls, Rasanen:2017, Enckell:2018a} and hillclimbing inflation \cite{Jinno:2017a, Jinno:2017b}.

On the gravity side, there are several different formulations of GR \cite{Einstein:1925, Einstein:1928a, Einstein:1928b, Einstein:1930, Krssak:2018ywd, Hehl:1976, Hehl:1978, Papapetrou:1978, Hehl:1981, Percacci:1991, Rovelli:1991, Nester:1998, Percacci:2009, Krasnov:2017, Gielen:2018, BeltranJimenez:2017, ferraris1982, BeltranJimenez:2019acz, Percacci:2020bzf}, which are equivalent for the minimal gravity action and minimally coupled matter, but differ for more complicated actions. The Higgs non-minimal coupling that is central to Higgs inflation breaks the equivalence between these formulations of GR and makes them physically distinct. The most studied alternative to the metric formulation in the context of Higgs inflation is the Palatini formulation \cite{Einstein:1925, ferraris1982}, where the metric and the connection are independent degrees of freedom \cite{Bauer:2008, Bauer:2010, Rasanen:2017, Enckell:2018a, Markkanen:2017, Rasanen:2018a, Rasanen:2018b, Racioppi:2018zoy, Rubio:2019, Racioppi:2019jsp, Shaposhnikov:2020fdv, Tenkanen:2020dge, Karananas:2020qkp, Langvik:2020, Shaposhnikov:2020frq, Gialamas:2020vto}.

An interesting motivation for the Palatini formulation is provided by the unitarity problem. It is a weak point of Higgs inflation that the non-minimal coupling spoils the SM cancellation between the EW gauge bosons and the Higgs in the scattering amplitudes. This leads to violation of tree-level unitarity around the EW vacuum at the scale $\Mpl/\xi$ in the metric formulation of GR \cite{Burgess:2009, Burgess:2010zq, Lerner:2009na, Lerner:2010mq, Hertzberg:2010, Bezrukov:2010, Bezrukov:2011a, Calmet:2013, Weenink:2010, Lerner:2011it, Prokopec:2012, Xianyu:2013, Prokopec:2014, Ren:2014, Escriva:2016cwl, Fumagalli:2017cdo, Gorbunov:2018llf, Ema:2019}. As the inflationary scale is $\Mpl/\sqrt{\xi}$ and $\xi\gg1$, this seems to imply that there is a region between the EW vacuum and the inflationary regime where the model is not a consistent quantum theory at the perturbative level. However, the unitarity cutoff is field-dependent \cite{Bezrukov:2010, Bezrukov:2011a, Xianyu:2013, Ren:2014, George:2015nza, Bezrukov:2017dyv}. When expanded around the inflationary plateau, the cutoff is above the inflationary scale indicating that dynamics during inflation can be studied by treating the model as a perturbative effective theory (high-momentum modes produced during preheating could probe the regime above the cutoff \cite{Ema:2016, DeCross:2016, Sfakianakis:2018lzf, Hamada:2020kuy}). 

At the inflationary scale, the theory reduces to the chiral SM, which is under control due to approximate shift symmetry on the inflationary plateau. However, the behaviour between the EW scale and the inflationary scale depends on the physics that is assumed to restore unitarity (either perturbatively, non-perturbatively or with new degrees of freedom). The unknown threshold corrections due to this physics can be phenomenologically modelled by adding arbitrary jumps to the SM couplings at the unitarity violation scale \cite{Bezrukov:2014bra,Bezrukov:2014ipa}. In the Palatini formulation, violation of perturbative unitarity and inflation both happen around the scale $\Mpl/\sqrt{\xi}$, so the theory is more reliable than in the metric formulation \cite{Bauer:2010, McDonald:2020, Shaposhnikov:2020fdv}. The theory is still not guaranteed to be under control due to the vicinity of the unitarity violation scale, although in general the scale of new physics can be higher than the tree-level perturbative unitarity violation scale, see \eg \cite{Aydemir:2012nz}.

Higher order curvature terms can also have a large effect on the inflationary dynamics and restore perturbative unitarity, as well as change the RG running \cite{Barbon:2015, Salvio:2015kka, Salvio:2017oyf, Kaneda:2015jma, Calmet:2016fsr, Wang:2017fuy, Ema:2017rqn, Pi:2017gih, He:2018mgb, He:2018gyf, Gorbunov:2018llf, Ghilencea:2018rqg, Wang:2018, Enckell:2018c, Antoniadis:2018ywb, Gundhi:2018wyz, Antoniadis:2018yfq, He:2020ivk, Bezrukov:2020txg, Ema:2020evi, Hamada:2020kuy, Bezrukov:2020txg, Ema:2020zvg, He:2020qcb}. In the Palatini case not all such terms introduce new degrees of freedom, so they have a less drastic impact on both the classical potential and loop corrections, meaning the formulation is more robust to corrections \cite{Enckell:2018b}. Alternative formulations of GR introduce the possibility of new non-minimal couplings between the Higgs and the connection, which can qualitatively change the potential \cite{Rasanen:2018b, Raatikainen:2019, Langvik:2020, Shaposhnikov:2020frq}. (In addition the Higgs kinetic terms can be non-minimally coupled, also in the metric case \cite{Germani:2010gm, Germani:2010ux, Kamada:2010qe, Kamada:2012se, Kamada:2013bia, Germani:2014hqa, Escriva:2016cwl, Fumagalli:2017cdo, Sato:2017qau, Granda:2019wip, Sato:2020ghj, Fumagalli:2020ody, Gialamas:2020vto}.) We consider only the simplest Higgs inflation action, where the Higgs doublet has a non-minimal coupling $\xi$ to the Ricci scalar.

The SM and chiral SM RG equations augmented with threshold jumps at the unitarity breaking scale have been applied to Higgs inflation in the hilltop case \cite{Enckell:2018a} in both the metric and the Palatini formulation, and to features similar to an inflection point in the metric formulation \cite{Enckell:2016}. (See also \cite{Fumagalli:2020ody} for the New Higgs inflation case.) We extend these studies to the case when the RG running creates an inflection point or a similar feature, often generically called a critical point, in the Palatini formulation. We study the effect of the choice of renormalisation scale. We check whether the slow-roll approximation is valid in the vicinity of the feature, or whether the field enters ultra-slow-roll.

In \sec{sec:calc} we set up the potential, explain how we treat the RG running, scan the parameter space and find the predicted range of the inflationary observables. In \sec{sec:disc} we discuss our results and compare to previous work, and in \sec{sec:conc} we summarise our findings.

\section{Critical point inflation} \label{sec:calc}

\subsection{Classical potential} \label{sec: class}

The Lagrangian of the SM coupled non-minimally to the Ricci scalar is (not writing explicitly the non-radial part of the Higgs, which enters only via its effect on RG running)
\begin{equation} \label{eq:higgs_SM_action}
	S = \int d^4 x \sqrt{-g} \qty[ \frac{1}{2}\left(M^2 + \xi h^2\right) g^{\a\b} R_{\a\b}(\Gamma, \pat\Gamma) - \frac{1}{2} g^{\a\b} \partial_\a h \partial_\b h - V(h) + \mathcal{L}_{\textrm{SM}} ] \ ,
\end{equation}
where $g_{\a\b}$ is the metric, $R_{\a\b}$ is the Ricci tensor, $M$ is a mass scale, $h$ is the radial Higgs field, $\xi$ is the non-minimal coupling, $V(h) = \frac{\lambda}{4} (h^2-v^2)^2$, where $v$ is the Higgs EW vacuum expectation value, and $\mathcal{L}_{\textrm{SM}}$ contains the rest of the SM. The Ricci tensor is built from the connection $\Gamma^\c_{\a\b}$, and as we consider the Palatini formulation, it is independent of the metric.

We make a conformal transformation to the Einstein frame and define the new scalar field $\chi$ with minimal coupling to gravity and canonical kinetic term \cite{Bezrukov:2007, Bauer:2008}:
\begin{equation} \label{eq:weyl_transform}
	g_{\alpha\beta} \rightarrow (1+\xi h^2)^{-1} g_{\alpha\beta} \ , \qquad \frac{\rmd \chi}{\rmd h} = \frac{1}{\sqrt{1 + \xi h^2}} \ ,
\end{equation}
where we have set  $M=\Mpl=1$ (see \cite{Rasanen:2018b} for discussion of this choice). We therefore have, fixing $\chi(h=0)=0$,
\bea
  h &=& \frac{1}{\sqrt{\xi}} \sinh(\sqrt{\xi} \chi) \ .
\eea
The SM Higgs potential in the original Jordan frame is
\bea
  V(h) &=& \frac{\l}{4} (h^2-v^2)^4 \approx \frac{\l}{4} h^4 \ ,
\eea
where we have taken into account that in the inflationary region $h\gg v$. In the Einstein frame, the potential is (again neglecting $v$) 
\bea
\label{eq:Uchi}
  U[h(\chi)] &=& \frac{V(h)}{(1 + \xi h^2)^2}
  = \frac{\l}{4} \frac{h^4}{(1 + \xi h^2)^2} \el
  &=& \frac{\l}{4\xi^2} \tanh^4(\sqrt{\xi} \chi)
  \equiv \frac{\l}{4} F(\chi)^4 \ ,
\eea
where we have defined $F(\chi)\equiv \frac{1}{\sqrt{\xi}}\tanh(\sqrt{\xi} \chi)$.
It is convenient to define the quantity \cite{Fumagalli:2016lls}
\begin{equation} \label{eq:delta}
	\delta \equiv \frac{1}{\xi h^2} \ ,
\end{equation}
which is small on the inflationary plateau.

\subsection{Quantum corrections}  \label{sec:quantum}

We take into account quantum corrections to the potential in the same way as in \cite{Enckell:2018a}, although our choice of renormalisation scale is different. For $\delta\ll1$ the theory is approximated by the chiral SM, and the one-loop correction to the Einstein frame effective potential in the $\overline{MS}$ scheme takes the form \cite{Bezrukov:2009db} 
\begin{equation} \label{eq:one_loop_corr}
	U_{\mathrm{1-loop}} = \frac{6m_W^4}{64\pi^2}\left(\ln \frac{m_W^2}{\mu^2} - \frac{5}{6} \right) +
				\frac{3m_Z^4}{64\pi^2}\left(\ln \frac{m_Z^2}{\mu^2} - \frac{5}{6} \right) -
				\frac{3m_t^4}{16\pi^2}\left(\ln \frac{m_t^2}{\mu^2} - \frac{3}{2} \right) \ .
\end{equation}
Here the squares of effective $W$, $Z$, and top quark masses are, respectively,
\begin{equation} \label{eq:large_masses}
	m_W^2 = \frac{g^2 F^2}{4} \ , \qquad m_Z^2 = \frac{\qty(g^2+g'^2) F^2}{4} \ , \qquad m_t^2 = \frac{y_t^2 F^2}{2} \ ,
\end{equation}
and other fermions are approximated to be massless. At one-loop order, the relevant chiral SM beta functions are given by  \cite{Bezrukov:2009db, Dutta:2007st}:
\bea \label{eq:large_betas}
	16\pi^2\beta_\lambda &=& -6y_t^4 + \frac{3}{8}\qty[2g^4 + (g'^2+g^2)^2] \ , \quad 16\pi^2\beta_{y_t} = y_t\qty(-\frac{17}{12}g'^2 - \frac{3}{2}g^2 - 8g_S^2 + 3y_t^2) \ , \el
	16\pi^2\beta_{g} &=& -\frac{13}{4}g^3 \ , \quad
	16\pi^2\beta_{g'} = \frac{27}{4}g'^3 \ , \quad
	16\pi^2\beta_{g_S} = -7g_S^3 \ .
\eea
In $\beta_\lambda$ we have omitted terms proportional to $\lambda g^2, \lambda g'{}^2$ and $\lambda y_t^2$ as they give subleading corrections in the limit $\lambda\simeq 0$ which is relevant for all cases with features. For the same reason, we neglect the running of the non-minimal coupling $\xi$. Rigorously, the theory reduces to the chiral SM only in the limit $\delta \rightarrow 0$; in the following we however assume the above treatment gives a decent approximation for the potential for all $\delta<1$.  

The leading order renormalisation group improved effective potential for $\delta<1$ is then
\beq
\label{U_RGimproved}
U(\chi) = \frac{\lambda(\mu)}{4} F(\chi)^4 + U_{\mathrm{1-loop}}(\chi,\mu)~,
\eeq
where $\lambda(\mu)$ denotes the solution of (\ref{eq:large_betas}) and $U_{\mathrm{1-loop}}(\chi,\mu)$ denotes (\ref{eq:one_loop_corr}) with all the couplings replaced by the corresponding solutions computed from (\ref{eq:large_betas}). The RG improved potential (\ref{U_RGimproved}) with the one-loop beta functions satisfies the Callan--Symanzik equation $\rmd U/\rmd{\rm ln}\mu = 0$ up to next-to-leading order terms (such as 
$\frac{m_i^4 \beta_j}{16 \pi^2 c_j}\ln\frac{m_i^2}{\mu^2}$ where $c_i$ denotes a coupling),
which would be cancelled by higher order loop corrections. To minimise the error from neglected higher order contributions, and the spurious dependence on $\mu$, the RG scale $\mu$ should be chosen to minimise the loop logarithms over the $\chi$ range of interest in the analysis. To do this, we choose $\mu$ equal to the largest mass scale involved in the loop corrections, following \eg \cite{Bezrukov:2014ipa,Enckell:2016},\footnote{This corresponds to prescription I according to the convention of \cite{Bezrukov:2008ej}, and is the preferred choice when computing the quantum corrections in the Einstein frame, as we do.}
\begin{equation} \label{eq:mu}
	\mu(\chi) = \frac{\gamma}{\sqrt{2}} F(\chi)~.
\end{equation}
Here $\c\approx0.34$, with the precise value chosen on a case by case basis so that $\sqrt{2}\mu$ is equal to the largest SM particle mass near the threshold scale $\delta = 1$. In \cite{Enckell:2018a, Rasanen:2018a} the scale $\mu$ was instead chosen so that the loop corrections vanish at the feature, whether it is a critical point, a local minimum or a hilltop. In principle that choice could have the drawback that, as different pieces of the loop correction have different signs, there could be accidental partial cancellations of large logarithms at the feature, leading to large corrections in its vicinity. We have checked how the results of our analysis change when using the prescription of \cite{Enckell:2018a, Rasanen:2018a}, \ie setting the loop correction to zero at the feature, instead of (\ref{eq:mu}). The difference in CMB predictions is negligible, showing that the analysis is robust against variation of the renormalisation scale. We have also varied the value of $\gamma$ in our prescription by a factor of $10$ up and down, and the difference is typically small, less than ten percent for the scalar perturbation amplitude $A_s$ and tensor-to-scalar ratio $r$, and smaller for the spectral index $n_s$. The differences are larger only on the edge of the parameter space, in particular on the small $r$ boundary in figure \ref{fig:areas} where the potential is very flat and sensitive to small corrections. Changing $\gamma$ moves this boundary slightly.

To connect the $\delta<1$ potential (\ref{U_RGimproved}) with low-energy physics, we match the chiral SM and SM couplings at the threshold scale $\mu_{\rm thres}$ that corresponds to $\delta = 1$, using the prescription (\ref{eq:mu}). As $\d=1$ corresponds to the transition between the SM region and the asymptotically flat plateau, this  ensures that the chiral SM RG equations are used only on the plateau.\footnote{As we have two scales in the RG corrected potential -- the field value and the renormalisation scale -- other criteria can lead to a situation where the SM RG equations extend up to the plateau or the chiral SM equations extend down from the plateau.}

For $\d>1$, \ie $\mu < \mu_{\rm thres}$, we run the couplings using the three-loop SM beta functions, and the central values of the observed strong coupling constant and particle masses \cite{Tanabashi:2019, Hoang:2020iah}
\begin{equation} \label{EWvalues}
	\frac{g_\textrm{S}^2(m_Z)}{4\pi}=0.1179\pm0.0010 \ , \quad m_H = 125.10 \pm 0.14\ \mathrm{GeV} \ , \quad m_t = 172.9\pm0.4\pm0.5\ \mathrm{GeV} \ ,
\end{equation}
where the second error in $m_t$ is an estimate of the systematic uncertainty in matching theory and experiment. The running is done with the code \cite{SMRunningCode}, with updated matching conditions between the observables and $\overline{MS}$ scheme parameters from \cite{Bednyakov:2015sca} together with input pole mass values $m_W=80.385$ GeV, $m_Z=91.1876$ GeV, and Fermi constant $G_F=1.1663787 \times 10^{-5}$ GeV\textsuperscript{$-2$}.

At the threshold scale, $\mu = \mu_{\rm thres}$,  we let the couplings $\lambda$ and $y_t$ jump by $\Delta\lambda$ and $\Delta y_t$ to parametrise the unknown effects of physics around the scale where perturbative unitarity is violated. For simplicity, we omit jumps in the gauge couplings and directly match their SM and chiral SM values at the threshold. We use the freedom in $\Delta\lambda$ and $\Delta y_t$ to create a feature in the quantum-corrected potential at scale $\delta=\delta_0<1$. This feature is a critical point, where the first and second derivatives of the potential are tuned to be small. We restrict our analysis to monotonically growing potentials; for potentials with a local maximum or minimum, see \cite{Enckell:2018a, Rasanen:2018a}.

\subsection{Equations of motion and inflationary observables} \label{sec:eom}

The background equations of motion in the Einstein frame read
\begin{equation} \label{background}
	3 H^2 = \frac{1}{2}\dot{\chi}^2 + U(\chi) \ , \qquad
	\ddot{\chi} + 3H\dot{\chi} + U'(\chi) = 0 \ ,
\end{equation}
where $H$ is the Hubble parameter, dot indicates derivative with respect to cosmic time $t$, and prime indicates derivative with respect to $\chi$. We solve these equations numerically for the quantum-corrected potential, starting in slow-roll far above the feature scale $\delta_0$.

The number of e-folds from the cosmic microwave background (CMB) pivot scale $k_*=0.05$ Mpc${}^{-1}$ to the end of inflation is
\begin{equation} \label{Nval}
  N = 56 - \frac{1}{4} \ln \frac{0.067}{r} \ ,
\end{equation}
where $r$ is the tensor-to-scalar ratio. This number depends on the duration of reheating. For the tree-level Higgs potential in the Palatini formulation, the reheating is practically instantaneous \cite{Rubio:2019}, and we have assumed that this is the case also here. We calculate $N$ numerically without using the slow-roll approximation.

To ensure approximate scale-invariance, we demand that the field is in slow-roll at the pivot scale, rejecting potentials where this is not the case. The CMB observables can then be expressed in terms of the slow-roll variables
\begin{equation}
	\epsilon \equiv \frac{1}{2} \qty(\frac{U'}{U})^2 \ ,
	\qquad
	\eta \equiv \frac{U''}{U} \ ,
	\qquad
	\zeta \equiv \frac{U'}{U} \frac{U'''}{U} \ ,
	\qquad
	\varpi \equiv \qty(\frac{U'}{U})^2 \frac{U''''}{U}  \ .
\end{equation}
The scalar amplitude, scalar spectral index, tensor-to-scalar ratio, running, and running of the running are, respectively,
\begin{equation} \label{eq:SR_observables}
\begin{split}
	A_s &= \frac{U}{24\pi^2 \epsilon} \ , \qquad\qquad\qquad\ n_s = 1-6\epsilon+2\eta \ , \qquad \qquad r = 16 \epsilon \ , \\
	\alpha_s &= 16\epsilon\eta - 24\epsilon^2 - 2\zeta \ , \qquad
	\beta_s = -192\epsilon^3 + 192\epsilon^2\eta - 32\epsilon\eta^2 - 24\epsilon\zeta + 2\eta\zeta + 2\varpi \ .
\end{split}
\end{equation}

The measured values of the CMB observables at the pivot scale 0.05 Mpc$^{-1}$ from Planck and BICEP2/Keck data are \cite{Planck2018}
\begin{equation} \label{eq:CMB_values}
\begin{split}
  A_s &= 2.099 \times 10^{-9} \, ,
  \qquad
  n_s = 0.9625 \pm 0.0048 \, ,
  \qquad
  r < 0.067 \, , \\
  \a_s &= 0.002 \pm 0.010 \, ,
  \qquad
  \b_s = 0.010 \pm 0.013 \, .
\end{split}
\end{equation}
We neglect the small observational uncertainty in $A_s$. We quote error bars as 68\% ranges and upper and lower limits as 95\% ranges. The limit on $r$ assumes zero running of the running.

\subsection{Numerical scan} \label{sec:numerics}

We have three free parameters: $\xi$, $\Delta\lambda$, and $\Delta y_t$. We fix them by the three conditions that the amplitude of scalar perturbations $A_s$ matches the observed value \re{eq:CMB_values} at the pivot scale, and the first two slow-roll parameters take the given values $\epsilon_0$ and $\eta_0$ at the given feature scale $\delta_0$. The feature conditions determine $\lambda$ and $y_t$ at the feature scale through a set of equations that we solve numerically. We then run the couplings back to the threshold scale $\mu_\mathrm{thres}$ to determine the jumps $\Delta\lambda$ and $\Delta y_t$. The coupling $\xi$ is determined iteratively to fix $A_s$ to its observed value \eqref{eq:CMB_values}.

We scan over all possible potentials by varying $\epsilon_0$, $\eta_0$, and $\delta_0$.\footnote{In principle, we have only two free parameters after $\xi$ has been traded for $A_s$, but numerically it is more convenient to vary three. We could cover all of the same possibilities if we, say, kept $\d_0$ fixed and varied $\epsilon_0$ and $\eta_0$ in a wider range, but such a scan would be less sensitive to strong features away from the point $\d_0$.} The slow-roll parameter $\eta$ is mostly negative on the inflationary plateau, but may be positive in a narrow range of field values near a strong feature. Hence, we do the scan in two parts. We first consider potentials for which $\eta > 0$ at some point on the plateau, and then potentials where $\eta < 0$ everywhere. For every potential, we solve the background evolution \eqref{background} numerically and calculate the CMB observables in the slow-roll approximation \eqref{eq:SR_observables}.

If $\eta>0$ for some field value, then there exists a field value where it crosses zero. We center on the zero-crossing point by setting $\eta_0=0$. We then perform a two-dimensional scan over $\epsilon_0$ and $\delta_0$, using $A_s$ to fix $\xi$. These potentials produce the strongest features and largest deviations from the tree-level result.

If $\eta < 0$ everywhere, we have $\eta_0 < 0$. We choose $\delta_0$ and perform a two-dimensional scan over $\epsilon_0$ and $\eta_0$, demanding $\eta < 0$ at all points, and again fixing $\xi$ with $A_s$. We repeat this for a number of $\delta_0$ values to ensure full coverage of the parameter space.

In all scans, we extend the parameter range in all directions as far as possible. The parameter space is limited at small $\delta_0$ and large $\epsilon_0$ by the fact that no value of $\xi$ can give the correct amplitude $A_s$. At large $\delta_0$, we demand that the feature is in the plateau region, $\delta_0 < 1$. Further limitations are set by our requirement that the potential is monotonic.

\begin{figure}
\begin{center}
	\includegraphics[scale=1]{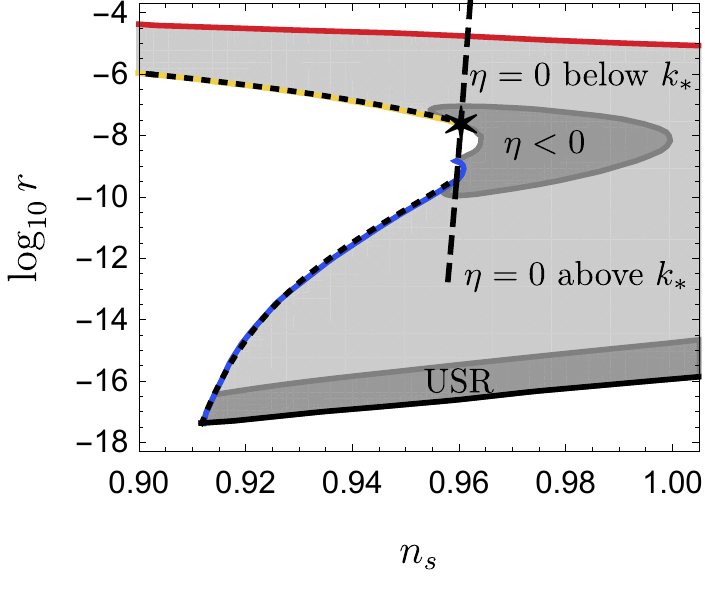}
	\hspace*{-5pt}
	\includegraphics[scale=1]{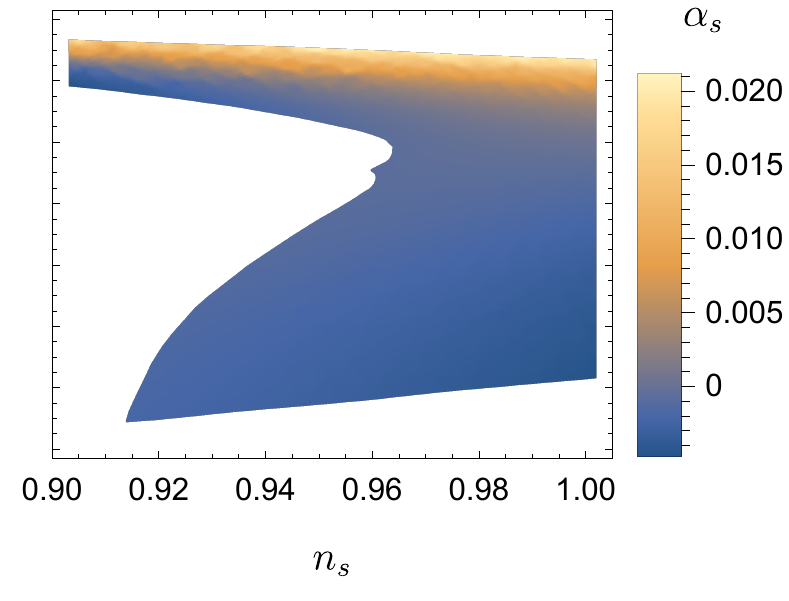}
\end{center}
\caption{Allowed points on the $(n_s, r)$-plane with different regions of the parameter space marked (left) and values of $\alpha_s$ shown (right), see explanation in the text. The allowed region continues to larger $n_s$ values on the right hand side and smaller $n_s$ values in the upper left corner. The dashed central line indicates the tree-level predictions with \eqref{Nval}, \eqref{eq:treelevel_CMB} and \eqref{eq:N_corrected}, and requiring $\lambda < 1$ for perturbativity. At the point marked with a star, the potential does not have a feature and our results merge with the tree-level results not only for $n_s$ and $r$ but also for $\alpha_s$ and $\beta_s$; see discussion around equation  \eqref{eq:our_tree_level_results}.}
\label{fig:areas}
\end{figure}

In figure \ref{fig:areas}, we show the resulting possible range for observables on the $(n_s, r)$-plane with $\alpha_s$ colour coded and various regions highlighted. In the middle dark grey area, the second slow-roll parameter $\eta$ is always negative (second part of our scan) and there is no strong feature in the potential. Elsewhere, there is a feature where $\eta=0$ (first part of our scan), either above (low $r$ region) or below (high $r$ region) the CMB pivot scale.

The feature is strongest on the blue and yellow boundaries on the left. Along the yellow line, the feature is close to the end of inflation, and we have USR near it. The field also enters USR in the lower dark grey area, in this case above the CMB pivot scale. In other regions, slow-roll applies everywhere. The CMB region and the feature region are well-separated in most of the parameter space, but start to overlap near the upper red line and the lower USR region. The viable inflationary region is limited from below by the black boundary, where the USR regime reaches the pivot scale and the spectrum ceases to be nearly scale-invariant. From above, the viable region is limited by the red boundary that marks the line where $\alpha_s = 0.022$, the 2$\sigma$ limit from \eqref{eq:CMB_values}. The dotted boundary lines correspond to the analytical approximations \eqref{eq:large_r_approx} and \eqref{eq:small_r_approx}, discussed in the next section. The wedge around $\log_{10} r = -8.8$ is due to the fact that for points right above the blue edge the potential is not monotonic; instead a local maximum and an adjacent minimum are formed around the feature scale. If such a potential supports successful inflation, it will be of the hilltop type, covered for the Higgs case in \cite{Enckell:2018a}.

Our range of $n_s$ is much wider than the observationally allowed range \re{eq:CMB_values}. If we restrict $n_s$ to the observed 2$\sigma$ range, we get $\alpha_s > -3.5 \times 10^{-3}$, and the upper values reach the $2\sigma$ upper limit $\alpha_s=0.022$ at the large $r$ edge of figure \ref{fig:areas}. Demanding both $n_s$ and $\alpha_s$ to be within the observed $2\sigma$ range, we get $-7.2 \times10^{-5} < \b_s < 6.1 \times 10^{-3}$ and $2.2 \times 10^{-17} < r < 2.0 \times 10^{-5}$. For these points, the non-minimal coupling is $1.1\times10^3<\xi<1.1\times10^9$. These values of the running of the running and the tensor-to-scalar ratio are well within the observational bounds \re{eq:CMB_values}. The value of $r$ can be larger than the typical tree-level Palatini result, but remains clearly below the tree-level metric result $r\approx4\times10^{-3}$ \cite{Bezrukov:2007}. The minimum value of $r$ is three orders of magnitude smaller than the smallest value in the Palatini tree-level case, $r=8\times10^{-14}$.

\subsection{Analytical approximations} \label{sec:analytics}

Let us try to understand the range of numerical predictions analytically. At tree level, the slow-roll approximation of section \ref{sec:eom} gives \cite{Bauer:2008}
\bea \label{eq:treelevel_CMB}
	n_s &=& 1 - \frac{2}{N_{\mathrm{sr}}} \approx 0.96 \  ,
	\quad \quad r = \frac{2}{\xi N^2_{\mathrm{sr}}} \approx 8 \times 10^{-4} \  \xi^{-1} \  , \el
	\alpha_s &=& -\frac{2}{N^2_{\mathrm{sr}}} \approx 8 \times 10^{-4} \  , \quad \beta_s = -\frac{4}{N^3_{\mathrm{sr}}} \approx 3 \times 10^{-5} \  , \quad
	\xi = \frac{\lambda N^2_{\mathrm{sr}}}{12\pi^2 A_s} \approx 10^{10} \lambda \  ,
\eea
where we have approximated $N_{\mathrm{sr}}\approx50$ at the pivot scale and used the observed value \eqref{eq:CMB_values} of $A_s$. Note that $r$ is suppressed by large values of $\xi$ as $r\propto1/\xi$. Here $N_{\mathrm{sr}}$ is given by the slow-roll approximation (also approximating that the value of $\chi$ at the end of inflation is small)
\begin{equation} \label{eq:N_SR}
	N_{\mathrm{sr}} \equiv \int_0^\chi \frac{d\tilde{\chi}}{\sqrt{2\epsilon(\tilde{\chi})}} \ .
\end{equation}
Due to the inaccuracy of the slow-roll approximation during the last e-folds of inflation, this deviates somewhat from the true number of e-folds $N$ in \eqref{Nval}; numerically \cite{Rubio:2019},
\begin{equation} \label{eq:N_corrected}
	N \approx N_{\mathrm{sr}} + 1.8 \, .
\end{equation}
This correction is of the same order as the one due to the dependence of the energy scale of inflation on $\xi$: the number $N=50$ corresponds to $\xi=10^9$, and the number of e-folds varies as $\Delta N=-\frac{1}{4}\ln(\xi/10^9)$. At tree level, if we take $\l=10^{-6}\ldots1$, then \re{eq:treelevel_CMB} gives $\xi=10^4\ldots10^{10}$, so $\Delta N=-0.6\ldots2.9$.

To obtain some analytical understanding of the RG improved potential, we rewrite the effective potential (\ref{U_RGimproved}) (with $\mu(\chi)$ determined by (\ref{eq:mu})) as 
\begin{equation} \label{eq:U_lambda_eff}
	U(\chi) = \frac{\lambda_\mathrm{eff}(\chi)}{4}F(\chi)^4 \ ,
\end{equation}
where the effective coupling $\lambda_\mathrm{eff}$ includes both the effect of the running of the coupling $\lambda$ as well as the loop correction terms. The amplitude of the scalar perturbation in \eqref{eq:SR_observables} gives
\begin{equation} \label{eq:As_corrected}
	A_s = \frac{\lambda_\mathrm{eff}}{6\pi^2 \xi^2 r}\ .
\end{equation}
We want to eliminate $\lambda_\mathrm{eff}$ to arrive at a relation between $r$ and $\xi$. In the potentials that have a strong feature, the effective coupling has to be of the same order as its running, which can be parametrically estimated from one of the contributions in  \eqref{eq:one_loop_corr}:
\begin{equation} \label{eq:lambda_eff_size}
	\lambda_\mathrm{eff} \sim -\frac{d \lambda_\mathrm{eff}}{d \ln \mu} \sim -\frac{d}{d \ln \mu} \left(\frac{3g^4}{128\pi^2}\right) = -\frac{12g^3 \beta_g}{128\pi^2} = \frac{39 g^6}{2048\pi^4} \sim 10^{-6} \  ,
\end{equation}
where we used the beta function \eqref{eq:large_betas} and set $g \approx 0.5$, a typical value at the inflationary scale according to our numerical results. Indeed, the numerics give $\lambda_\mathrm{eff} \approx 3.1 \times 10^{-6}$ on average, with only small variation around this. Combining this with equation \eqref{eq:As_corrected} gives
\begin{equation} \label{eq:xi_r}
	r \approx \frac{25}{\xi^2} \ .
\end{equation}
This is accurate for most of the parameter space; in some regions the prefactor varies between 5 and 31. The stronger suppression $r\sim10/\xi^2$ explains the small values of $r$ compared to the tree-level case, where $r\sim10^{-3}/\xi$.

Substituting \re{eq:xi_r} into the tree-level results \eqref{eq:treelevel_CMB} and using the adjusted value of $N$ \eqref{eq:N_corrected},  we find the only point where the results of our numerical analysis with a strong feature coincide with the predictions of a tree-level potential for all CMB observables:
\bea \label{eq:our_tree_level_results}
	n_s &\approx& 0.96 \ , \qquad\qquad\qquad r \approx 2.5\times 10^{-8} \  , \quad\quad\ \ N \approx 52 \  , \el
	\alpha_s &\approx& -7.8 \times 10^{-4} \  , \qquad \beta_s \approx -3.1 \times 10^{-5} \  , \qquad \xi \approx 3.2\times10^4 \ .
\eea
This model is marked with a star in figure \ref{fig:areas}. The other tree-level results with a different value of $\xi$, marked with a dashed line, are of course still possible when the quantum corrections are included, but our numerical analysis, which is restricted to the fine-tuned cases with strong features, does not see them.

Armed with the relation \eqref{eq:xi_r}, we can make further analytical approximations on the edges of the parameter space shown in figure \ref{fig:areas}. This is true particularly along the yellow and blue curves, which correspond to a strong feature with $\epsilon\approx\eta\approx0$ at the critical point.

On the yellow line, the critical point is below the CMB pivot scale and the slow-roll approximation fails in its vicinity. The field enters a regime of ultra-slow-roll (USR) \cite{Faraoni:2000vg, Kinney:2005vj, Martin:2012pe, Dimopoulos:2017ged, Pattison:2018}. Above the feature, the evolution is similar to the tree-level case. The effect of the USR period on the spectrum on the CMB scales is to reduce the number of e-folds spent on the plateau compared to the tree-level case \eqref{eq:treelevel_CMB}, so we have
\begin{equation} \label{eq:CMB_with_Neff}
	n_s = 1 - \frac{2}{N_\mathrm{eff}} \  , \qquad
	r = \frac{2}{\xi N^2_\mathrm{eff}} \  , \qquad
	\alpha_s = -\frac{2}{N_\mathrm{eff}^2} \  , \qquad
	\alpha_s = -\frac{4}{N_\mathrm{eff}^3} \  ,
\end{equation}
where $N_\mathrm{eff}$ varies between zero and $50$. Combining this result with \eqref{eq:xi_r} gives
\begin{equation} \label{eq:large_r_approx}
	r = \qty(\frac{1-n_s}{3.1})^4 \  , \qquad
	\alpha_s = -\frac{1}{2}\qty(1-n_s)^2 \  , \qquad
	\beta_s = -\frac{1}{2}\qty(1-n_s)^3 \  .
\end{equation}
This approximation is marked with a dotted line in figure \ref{fig:areas} and agrees well with the yellow boundary given by the numerical scan.

On the blue line, the critical point is above the CMB pivot scale. We expand the effective coupling $\lambda_\mathrm{eff}$ in the potential \eqref{eq:U_lambda_eff} up to second order in its running, fixing the coefficients so that $\epsilon\approx\eta\approx 0$ for $\delta=\delta_0$. This gives
\begin{equation} \label{eq:lambda_eff_expansion}
	\lambda_\mathrm{eff}(\delta) = \lambda_0 - 4\lambda_0 \ln \frac{F(\delta)}{F(\delta_0)} + 8\lambda_0 \ln^2 \frac{F(\delta)}{F(\delta_0)} \  ,
\end{equation}
where $F(\delta) = \frac{1}{\sqrt{\xi(1+\delta)}}$ as before. Expanding everything to leading order in the small parameters $\delta$ and $\delta_0$, we get in the slow-roll approximation
\bea \label{eq:small_r_approx}
	N_{\mathrm{sr}} &=& \frac{\delta_0(2\delta-\delta_0) + 2\delta(\delta-\delta_0)\ln\qty(1-\frac{\delta_0}{\delta})}{16\xi \delta_0^3\delta(\delta-\delta_0)} \, , \el
	n_s &=& 1 - 32\xi\delta(\delta-\delta_0)(3\delta-\delta_0) \, , \el
	r &=& 512\xi\delta^2(\delta-\delta_0)^4 \  .
\eea
Combining this with the value \eqref{eq:xi_r} for $r$ and the approximation $N \approx N_{\mathrm{sr}} + 2.5$ (note that the optimal shift in $N$ is different from the tree-level case \eqref{eq:N_corrected}), the result, again marked with a dotted line, agrees with the numerical results on the blue line in figure \ref{fig:areas}. This approximation also works for $\alpha_s$ and $\beta_s$, but we omit the lengthy expressions which contain powers up to six and nine in $\delta$ and $\d_0$.

This approximation helps us understand the lowest allowed $r$ values in figure \ref{fig:areas}. Varying $\delta$ for a fixed potential (fixed $\delta_0$ and $\xi$), the maximum value of $\eta$ turns out to be
\begin{equation} \label{eq:eta_max}
	\eta_\mathrm{max} = \frac{32}{243}(10 + 7 \sqrt{7}) \delta_0^3 \xi \, ,
\end{equation}
located somewhat above the CMB pivot scale. This value grows with $\xi$ for small $r$ values, until it reaches unity at $\log_{10} r \approx -16.8$. Around this scale, slow-roll is replaced by USR which for slightly smaller $r$ reaches the pivot scale and ruins the scale-invariance of the CMB spectrum. This corresponds to the bottom black boundary in figure \ref{fig:areas}.

\subsection{Varying the EW scale parameters} \label{sec:EW}

\begin{figure}
\begin{center}
	\includegraphics[scale=1]{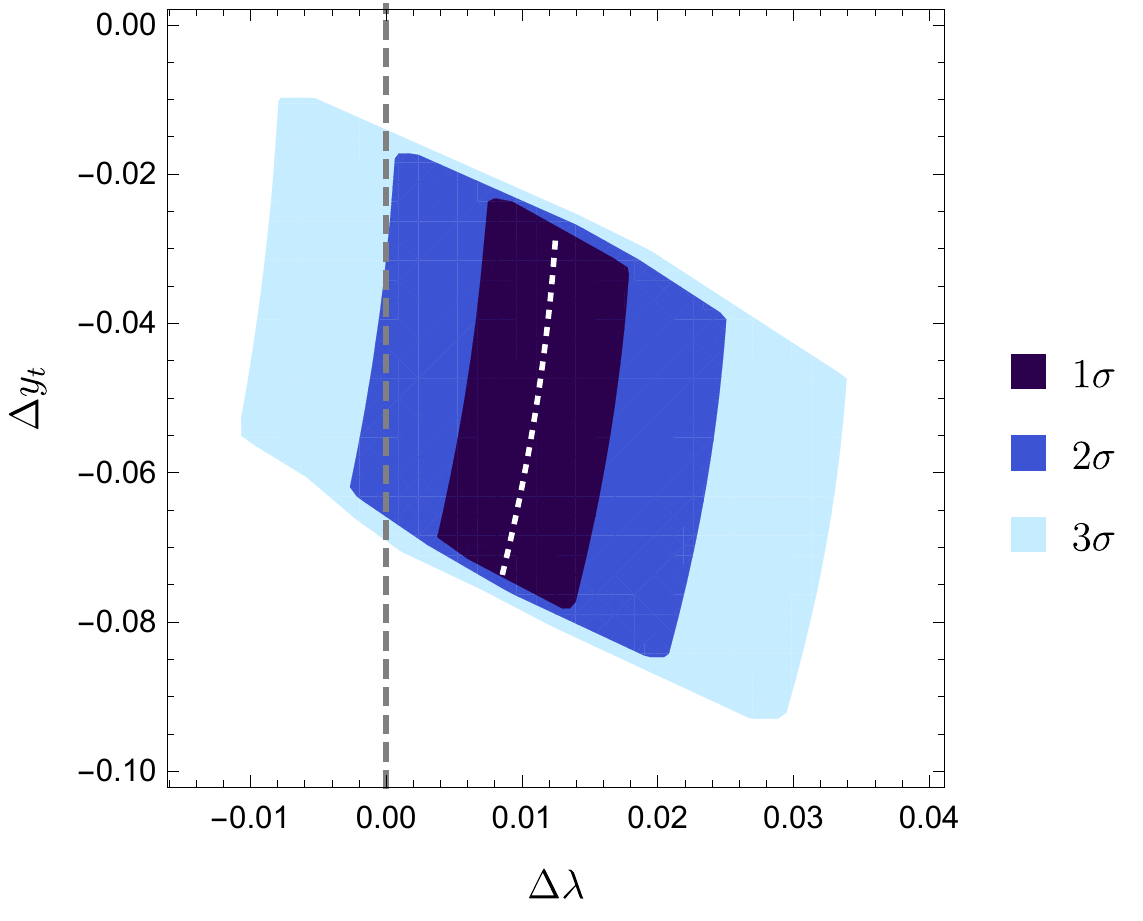}
\end{center}
\caption{Values of the jumps $\Delta \lambda$ and $\Delta y_t$ that lead to a potential with a feature but with the CMB observables within the 2$\sigma$ limits \eqref{eq:CMB_values}. The dashed white line corresponds to the central EW parameter values \eqref{EWvalues}. In the coloured regions the SM parameters vary within their $1\sigma$, $2\sigma$, and $3\sigma$ limits.}
\label{fig:sigma}
\end{figure}

In the above analysis we have kept the SM EW scale input parameters \re{EWvalues} fixed to their central values. The running of the Higgs quartic coupling is well known to be sensitive to changes in these parameters. This sensitivity does not transfer to inflationary observables, because their variation is degenerate with changes in the jumps $\Delta \lambda$ and $\Delta y_t$. Still, varying the EW parameters does affect field evolution after inflation, as well as the size of the jumps needed to get a potential with strong features.

For central values of the EW parameters,  the $\lambda$ jumps span the range $\Delta \lambda = 0.009 \dots 0.012$, as shown in figure \ref{fig:sigma}. The $y_t$ jumps are negative, $\Delta y_t = -0.07 \dots {-0.03}$, and bring $y_t$ down to about $0.40$. Note that the values of the jumps depend on the choice of the threshold scale and the renormalisation scale. For the central values,  all cases with features are such that $\lambda$ runs negative below the threshold scale, and $\Delta \lambda > 0$ raises it to a small but still negative value in the inflationary regime (the added quantum corrections make the potential positive). 

When we vary the EW scale input parameters \re{EWvalues} within the 2$\sigma$ range, the ranges of the jumps change to $\Delta \lambda = -0.003 \dots 0.025$ and $\Delta y_t = -0.085 \dots {-0.017}$, as shown in figure \ref{fig:sigma}. Changing the top quark mass has the largest effect. The cases $\Delta \lambda < 0$ in the figure are such that $\lambda$ stays positive all the way to the threshold scale, and they correspond to top mass values below the central value.  The figure shows that cases with $\Delta y_t$ = 0 do not occur within 3$\sigma$ limits of EW parameters. In particular, this means that there are no viable inflationary models with a feature and  both jumps set to zero. For zero jumps, the model either exhibits an instability at the inflationary scale (the effective potential is negative) or gives the tree-level results \eqref{eq:treelevel_CMB}.\footnote{In a previous study \cite{Enckell:2018a}, a viable hilltop scenario with zero jumps was found within the 2$\sigma$ range, but the spectral index $n_s=0.94$ is now in strong tension with updated observations. The threshold and renormalisation scales were also chosen differently there.} 

Close to the $\Delta \lambda = 0$ line, there are cases where a local minimum is formed at the threshold scale. (The minimum could also be formed above the threshold scale, but we exclude such cases in our analysis.) It would be interesting to investigate reheating in detail in such a setting. The effect of a second minimum below the inflationary region depends on the detailed form of the potential, as the field may roll fast enough to pass over the second minimum, or the minimum may be uplifted by thermal corrections during reheating \cite{Bezrukov:2014ipa, Rubio:2015zia, Enckell:2016}.

\section{Discussion} \label{sec:disc}

\subsection{Comparison to previous work}

In \cite{Rasanen:2017}, Higgs inflation with a critical point was studied in both the metric and the Palatini formulation with a simple analytic approximation for the potential, similar to \re{eq:lambda_eff_expansion}, but without the linear logarithmic term, and assuming slow-roll. We consider a wider range of possibilities, and find a larger range of values for the spectral index, in particular we can have $n_s>0.96$. Our upper range for $r$ is $2\times10^{-5}$, compared to $7\times10^{-5}$ in \cite{Rasanen:2017}. Our ranges for $\a_s$ and $\b_s$ (when conditioned the same way) agree to within a factor of about 2 with the results of \cite{Rasanen:2017}. The maximum value of $\xi$ found there was $6\times10^6$, so the analysis missed the possible strong features for large values of $\xi$ up to $1.1\times10^9$ that we find.

If we compare to metric case Higgs inflation with a critical point studied with the RG improved potential in \cite{Enckell:2016}, the spectral index in both cases covers a much wider range than the observationally allowed window, and the possible range of values for $\a_s$ is roughly the same. In contrast, there is no overlap between the allowed $r$ values: in the Palatini case, $r$ is always smaller. This agrees with previous analyses showing that even when quantum corrections are taken into account, the metric and the Palatini formulation of Higgs inflation remain distinguishable \cite{Rasanen:2017, Enckell:2018a}. This is also true when a $R^2$ term is included in the action \cite{Enckell:2018b, Enckell:2018c}, although a full analysis including both the $R^2$ term and quantum corrections has not been done. (If new gravitational terms are included in the action in the Palatini case, $r$ can take values all the way to the observational upper limit \cite{Rasanen:2018b, Langvik:2020}.)

In \cite{Shaposhnikov:2020fdv} it was argued that the tree-level predictions of Higgs inflation in the Palatini case are robust to loop corrections. The threshold scale is roughly the same as ours. It was argued that new physics there is expected to lead at most to a jump of $\Delta\l=6\times10^{-4}$; the jump of $y_t$ was neglected. On this basis, it was argued that the tree-level predictions for $n_s$ and $r$ are unchanged, and that the non-minimal coupling is bounded as $1.0\times10^6<\xi<6.8\times10^7$, which implies $1.1\times10^{-11}<r<0.8\times10^{-9}$. For the central values of the EW parameters, we find that $\Delta\l$ has to be at least $0.009$ in order to produce a strong feature. This agrees with the conclusion of \cite{Shaposhnikov:2020fdv} that $\Delta\l=6\times10^{-4}$ is not enough to significantly change the inflationary predictions. However, if we move away from the mean values, the situation changes. As discussed in \sec{sec:EW} and shown in \fig{fig:sigma}, we can have a strong feature even when $\Delta\l=0$ (if $\Delta y_t=-0.03$) when we allow the EW parameters to vary within their $2\sigma$ ranges.

We conclude that the predictions of Higgs inflation in the Palatini case are not tightly fixed by the tree-level action in the presence of threshold corrections. However, generating a strong feature requires the threshold and loop corrections to be highly tuned, so in a typical case there are no features and the tree-level results are not significantly modified.

\subsection{Unitarity}

We have used the RG improved effective potential, and chosen the renormalisation scale to be a field-dependent function $\mu(\chi)$ that minimises the loop corrections. This follows the procedure of particle physics scattering analysis, where the renormalisation scale is chosen to minimise the corrections at the experimental scattering energy. We have then defined the threshold scale $\mu_{\rm thres}$ as the scale that corresponds to $h\sim\chi\sim1/\sqrt{\xi}$, the cutoff of the effective theory expanded around the EW scale, and matched the chiral SM and SM couplings at this scale up to jumps in $\lambda$ and $y_t$. This choice of the threshold scale is customary and coincides with the energy scale where tree-level unitarity is violated in scattering processes involving the Higgs doublet, the $W$ and $Z$ bosons (and, in the Jordan frame, the graviton), computed by expanding around the EW vacuum \cite{Burgess:2009, Bauer:2010, Xianyu:2013, Ren:2014} (see also \cite{Burgess:2010zq, Lerner:2009na, Lerner:2010mq, Hertzberg:2010, Bezrukov:2010, Bezrukov:2011a, Calmet:2013, Weenink:2010, Lerner:2011it, Prokopec:2012, Prokopec:2014, Escriva:2016cwl, Fumagalli:2017cdo, Gorbunov:2018llf, Ema:2019}). Let us comment on the role of such unitary violating scatterings for Higgs inflation. 
 
There are a few different scales involved: the renormalisation scale $\mu(\chi)$, the effective masses (\ref{eq:large_masses}), the field value $\chi$, the energy scale $U(\chi)^{1/4}$ of the potential, and the Hubble scale $H$, which determines the amplitude of the power spectrum of field perturbations. In the Palatini tree-level case, the field value $\chi\sim10/\sqrt{\xi}$ in the CMB region is close to the unitarity violation scale $\sim1/\sqrt{\xi}$. So is the energy scale of the potential $U^{1/4}\sim\l^{1/4}/\sqrt{\xi}$; the Hubble scale $H\sim\l^{1/2}/\xi$ is much smaller. In the tree-level case the original Jordan frame Higgs field value $h\approx20$ is orders of magnitude above the unitarity cutoff, unlike in the metric case where $h\sim10/\sqrt{\xi}$. In our numerical results, $\chi=(3\dots8)/\sqrt{\xi}$ for the points that agree with observations, corresponding to $h=(10\dots1148)/\sqrt{\xi}$. Values $h>1$ occur only close to the point marked with the star in \fig{fig:areas}; the largest value of $h$ is 7.5, smaller than in the tree-level case.

None of these scales is directly related to the tree-level unitarity violation energy obtained from scattering calculations. Such scattering processes are not expected to play any role during inflation when all particle numbers are exponentially diluted, and the relevant momenta and amplitudes of the field perturbations are determined by the Hubble scale, which is far below $1/\sqrt{\xi}$. So even if perturbative unitarity is violated for the scattering processes, meaning they cannot be reliably calculated using the theory in the inflationary regime, this does not necessarily imply that the theory cannot be used to reliably calculate the processes that are relevant for inflation. (The situation during preheating can be different if modes with momenta above the scattering unitarity violation scale are produced \cite{Ema:2016, DeCross:2016, Sfakianakis:2018lzf, Hamada:2020kuy}.) Nevertheless, going beyond the tree-level, the non-renormalisable terms in the potential (\ref{eq:Uchi}) become of order unity at $h\sim\chi\sim1/\sqrt{\xi}$, meaning that the effective theory ceases to be predictive because of the infinite hierarchy of unsuppressed loop corrections. So regardless of the tree-level unitarity issue, there is reason to doubt the validity of the theory close to this scale, and thus we add jumps to the couplings as effective threshold corrections. In contrast, in the inflationary regime the loop corrections are suppressed by the assumed quantum extension of the classical shift symmetry.

\subsection{Spectral distortions}

In the upper edge of figure \ref{fig:areas}, the running of the spectral index $\alpha_s$ and the running of the running $\beta_s$ are large and positive, implying growing amplitude of perturbations towards smaller scales and an enhanced $\mu$-distortion signal, possibly detectable by next generation spectral distortion experiments \cite{Chluba:2019kpb}. The relation between the $\mu$-distortion signal and the running $\alpha_s$ in single field models was studied in \cite{Dent:2012ne, Cabass:2016giw}, and extended to $\beta_s$ in \cite{Cabass:2016ldu}. These general results apply also to our setup, apart from configurations close to the red line in figure \ref{fig:areas}, where $\alpha_s=0.022$.  For these points, the strong feature in the potential leads to scale-dependence of $\PR(k)$ that is not fully described by $\alpha_s$ and $\beta_s$. Using \cite{Hu:1994bz,Chluba:2012gq,2012A&A...543A.136K}, and assuming instant transition between the $\mu$ and $y$ distortion at redshift $z = 5 \times 10^{4}$ as in \cite{Dent:2012ne,Cabass:2016giw,Cabass:2016ldu}, we calculate the $\mu$-signal using with the full numerical result for $\PR(k)$. On the edge where $\alpha_s=0.022$, we find $\mu \approx 9 \times 10^{-8}$ for the points where $n_s$ has the central value \eqref{eq:CMB_values}. This should be detectable with future surveys \cite{Chluba:2019kpb}. Using the expansion in terms of $n_s$, $\alpha_s$, and $\beta_s$ would instead give $\mu \approx 6 \times 10^{-8}$, so the violation of this approximation enhances the signal by $50\%$.

\section{Conclusions} \label{sec:conc}

\para{The predictions of critical point Palatini Higgs inflation.}

We have studied the range of predictions for Palatini Higgs inflation in the case when loop corrections generate a critical point in the potential, a feature similar to an inflection point. We have scanned over the parameter space formed by the non-minimal coupling $\xi$ and jumps in the quartic Higgs coupling $\l$ and top Yukawa coupling $y_t$.

The scan shows that the spectral index $n_s$ can take all values in the observationally allowed range, and beyond. Requiring $n_s$ to be within the observed $2\sigma$ range, its running is limited from below as $\a_s> -3.5\times10^{-3}$. Requiring also the running to be below the observed $2\sigma$ upper limit, the running of the running is $-7.2 \times10^{-5} < \b_s < 6.1 \times 10^{-3}$. The similarly conditioned tensor-to-scalar ratio is small, $2.2 \times 10^{-17} < r < 2.0 \times 10^{-5}$. The lower end is among the smallest tensor-to-scalar ratios of any inflationary model studied so far. The upper end is below the smallest value found in the metric formulation, even when loop corrections are taken into account. Notably, the Palatini tree-level relation $r=2/(\xi N^2)$ is replaced by the relation $r\sim10/\xi^2$ when we have a strong feature. The value of the non-minimal coupling is $1.1\times10^3<\xi<1.1\times10^9$, so it can be smaller than in the tree-level metric case where $\xi\approx10^4$, but not quite as small as the values $\xi\sim400$ found in the critical point metric case \cite{Enckell:2016} or $\xi=180$ in the hilltop metric case \cite{Enckell:2018a}.

In part of the parameter space the feature leads to a period of ultra-slow-roll. This can't happen too close to the CMB pivot scale, since it would lead to too large deviations from scale-invariance. If ultra-slow-roll occurs below the pivot scale, it reduces the number of e-folds spent on the plateau. CMB predictions are then given by the same equations as in the tree-level case, but with a smaller number of e-folds inserted into the equations. If ultra-slow-roll happens above the pivot scale, it is possible for the pivot scale to lie at a very flat point near the feature, leading to the extremely small $r$ values.

We have also studied the jumps in the couplings required to produce such strong features. For the measured central values of the SM EW scale parameters, we need $\Delta \lambda = 0.009 \dots 0.012$ and $\Delta y_t = -0.07 \dots {-0.03}$ for a feature to form. If we allow the SM parameters to vary within $2\sigma$, we can have a strong feature even with a continuous $\l$, but the jump of $y_t$ must be non-zero.

The CMB predictions of Higgs inflation can also be modified in other ways, e.g. by adding an $R^2$ term into the action. Similar to our results, this also widens the allowed parameter space \cite{Wang:2017fuy, Gundhi:2018wyz, Enckell:2018c}. However, in the case of loop corrections, most of the predictions are accessed only with highly tuned choices of parameters, unlike in Higgs-$R^2$ inflation.

We conclude that loop and threshold corrections can generate a feature with interesting phenomenology and significant observational effects in the Palatini formulation. Nevertheless, despite the dispersion of predictions both in the metric and in the Palatini case, for the simplest tree-level Higgs inflation action, the two formulations remain distinguishable due to lack of overlap in the predicted values of the tensor-to-scalar ratio.

\acknowledgments

This work was supported by the Estonian Research Council grants PRG803 and MOBTT5 and by the EU through the European Regional Development Fund CoE program TK133 ``The Dark Side of the Universe''.

\bibliographystyle{JHEP}
\bibliography{infl}

\end{document}